\documentclass[prl, showpacs, twocolumn]{revtex4}
\usepackage{graphicx}
\usepackage{times}
\usepackage{bm}
\usepackage{amsmath, verbatim}
\usepackage[varg]{txfonts}

\def\be{\begin{equation}}
\def\ee{\end{equation}}
\def\bea{\begin{eqnarray}}
\def\eea{\end{eqnarray}}
\def\bse{\begin{subequations}}
\def\ese{\end{subequations}}

\def\be{\begin{eqnarray}}
\def\ee{\end{eqnarray}}

\begin{document}

\title{Proximity effect at the superconductor - topological insulator interface}
\author{Tudor D. Stanescu$^{2,1}$}
\author{Jay D. Sau$^1$}
\author{Roman M. Lutchyn$^1$}
\author{S. Das Sarma$^1$}
\affiliation{
$^1$Condensed Matter Theory Center and Joint Quantum Institute, Department of Physics, University of
Maryland, College Park, Maryland 20742-4111, USA\\
$^2$
Department of Physics, West Virginia University, Morgantown, WV 26506, USA}

\begin{abstract}
We study the excitation spectrum of a topological insulator in contact with an $s$-wave superconductor, starting from a microscopic model, and develop an effective low-energy model for the proximity effect. In the vicinity of the Dirac cone vertex, the effective model describing the states localized at the interface is well approximated by a model of Dirac electrons experiencing superconducting $s$-wave pairing.
Away from the cone vertex, the induced pairing potential develops a $p$-wave component with a magnitude sensitive to the structure of the interface. Observing the induced $s$-wave superconductivity may require tuning the chemical potential close to the Dirac point. Furthermore, we find that the proximity of the superconductor leads to a significant renormalization of the original parameters of the effective model describing the surface states of a topological insulator.
\end{abstract}
\date{\today}


\maketitle


Topological insulators have emerged as an important class of materials characterized by robust quantum properties that can provide a significant potential for new device architectures. The recent observations of a two-dimensional (2D) topological insulator (TI) state in HgTe quantum wells~\cite{Konig2007} and of three-dimensional TI behavior in a family of bismuth-based materials, including Bi$_2$Se$_3$ and Bi$_2$Te$_3$~\cite{Hsieh2008,Hsieh2009,Roushan2009,TZhang2009,Xia2009}, represent a significant impetus to the experimental efforts in this area and a stimulus for the theoretical study of  model systems with topologically ordered ground states~\cite{Kitaev2003,KaneMele2005,Bernevig2006,Murakami2006,Wu2006,Fu2006,Fu2007}. 
The non-trivial topological properties of an insulator are intrinsically linked to the existence near the boundaries of gapless surface (or edge) states  that are robust against disorder and interaction and have a characteristic Dirac type spectrum.

Of special interest are the states that develop at the interface between a TI and an $s$-wave superconductor (SC), where the proximity effect generates a quasi-two-dimensional quantum state similar to a spinless $p_x+ip_y$ superconductor, but without breaking time-reversal symmetry. This state supports Majorana bound states at vortices and was recently proposed by Fu and Kane~\cite{Fu2008} as a platform for fault-tolerant topological quantum computation.
The proposal has generated a lot of interest due to the rather simple, weakly correlated nature of the underlying system. However, the existing calculations of the properties of this system are based on a phenomenological treatment of the proximity effect, where an $s$-wave pairing term is added to an effective Hamiltonian describing the TI surface. 
In addition, the general nature of the proximity effect between Dirac-type materials, such as graphene, and $s$-wave SCs has been a subject of great interest recently~\cite{Heersche'07,Miao'09,lutchyn'08}. 

In this Rapid Communication we study the proximity effect at the interface between a TI and an $s$-wave SC starting from a microscopic model. In particular, we address and clarify the following questions: (1) how is the low-energy spectrum of the TI modified by the proximity effect induced at the interface? What is the dependence of this spectrum on the microscopic parameters of the system and what is the role of dynamical effects? (2) What is the most general effective model for the interface and what is the dependence of its parameters on the relevant microscopic quantities?  The answers to these questions have direct consequences for the stability of the Majorana bound states that can be obtained at the interface.

To study the proximity effect between an $s$-wave SC and the surface states of a strong TI, we consider a minimal microscopic model defined by the following Hamiltonian:  
\begin{equation}
H_{\rm tot}=H_{\rm TI}+H_{\rm SC}+H_{\rm t}. \label{Htot}
\end{equation}
Here $H_{\rm TI}$ is the TI term given by the tight-binding model on a diamond lattice with spin-orbit interactions~\cite{Fu2007},
\begin{align}
H_{\rm TI}\!=\! t\!\sum_{\langle ij\rangle, \sigma}  c_{i\sigma}^\dagger c_{j\sigma}
\!+\! i\lambda_{SO}\! \sum_{\langle\langle ij \rangle\rangle}\!
 {\bf S}_{\sigma,\sigma'}\! \cdot \! ({\bf d}_{ij}^1 \! \times \! {\bf d}_{ij}^2) c_{i\sigma}^\dagger c_{j\sigma'}.
 \label{eq:tbmodel}
\end{align}
The first term is the nearest-neighbor hopping connecting the
two fcc sublattices of the diamond lattice.  The second term connects
second-order neighbors with a spin- and direction-dependent amplitudes. The direction dependence is given by the bond vectors ${\bf d}_{ij}^{1,2}$  traversed between sites $i$ and $j$. The superconducting term in Eq. (\ref{Htot}) is given by mean-field Hamiltonian
\begin{align}\label{eq:scmodel}
H_{\rm SC}= \sum_{\langle ij\rangle, \sigma} t_{ij}^{\prime} a_{i\sigma}^\dagger a_{j\sigma}+\sum_{i} (\Delta_0 a^\dag_{i\uparrow} a^\dag_{i\downarrow}+h.c.)
\end{align}
The model is defined on a simple hexagonal lattice with a lattice constant for the triangular basis that ensures simple interface matching conditions between the SC and the ($1,1,1$) surface of the TI. In general, the nearest neighbor hopping $t_{ij}^{\prime}$ has different values along directions parallel and perpendicular to the interface. The tunneling term in Eq.   (\ref{Htot}) is
\begin{align}
H_{\rm t}=\sum_{\langle ij \rangle }  \tilde t (a_{i\sigma}^\dagger c_{j\sigma}+c^\dag_{j\sigma} a_{i\sigma}), \label{Ht}
\end{align}
were $\tilde t$ is the tunneling matrix element that characterizes the transparency of the interface between TI and SC. The fermion operators  $c_{j\sigma}$ and $a_{i\sigma}$ operate in the Hilbert space of the TI and superconductor, respectively.  

\begin{figure}[tbp]
\begin{center}
\includegraphics[width=0.45\textwidth]{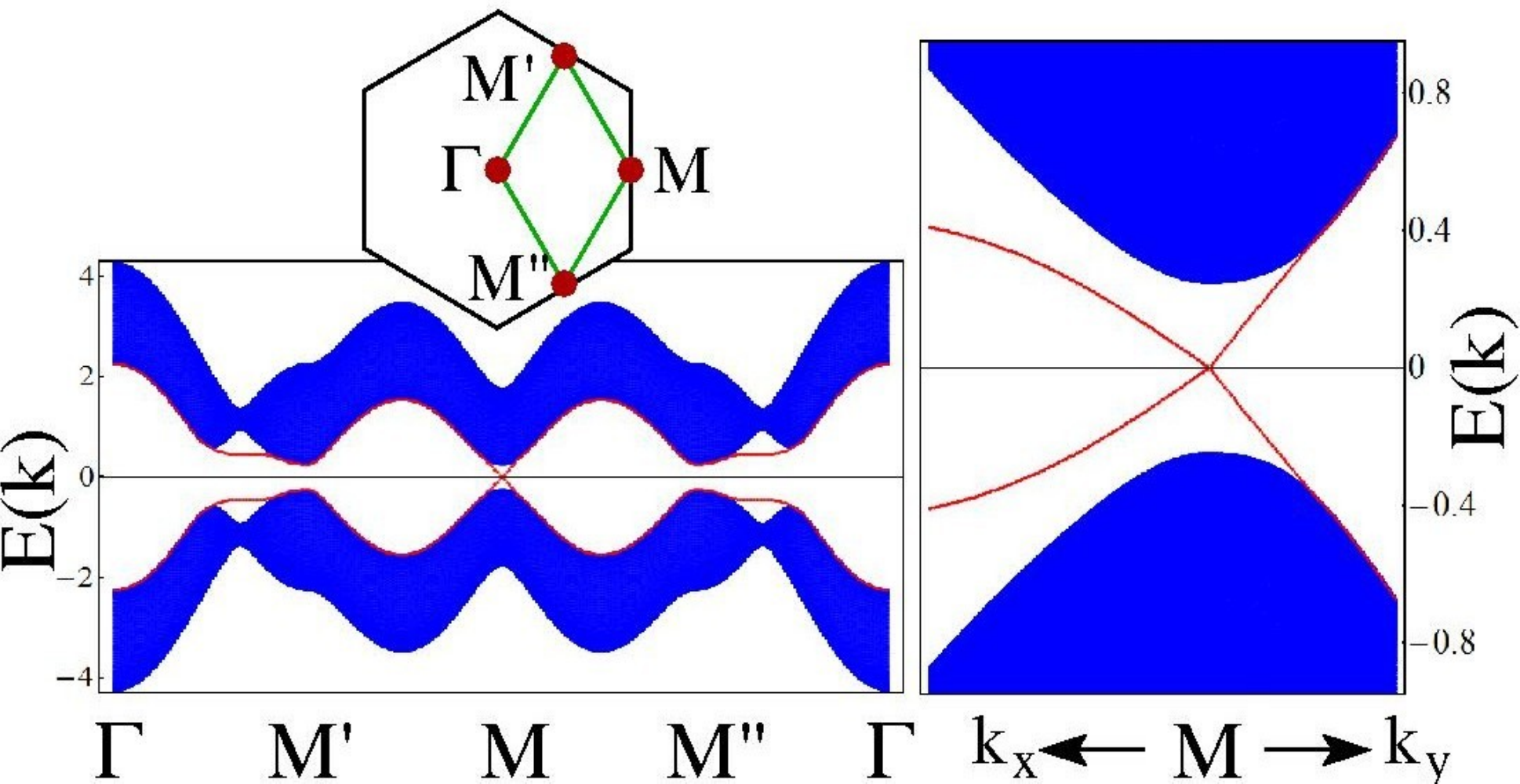}
\vspace{-3mm}
\end{center}
\caption{(Color online) Band structure for a slab with a ($1,1,1$) surface of a  model TI described by Eq. (\ref{eq:tbmodel}). The blue (dark gray) areas represent bulk states and the in-gap lines are surface states. The right panel shows a zoom-in of the anisotropic Dirac cone at the M point of the surface Brillouin zone.} \vspace{-4mm}
\label{Fig1}
\end{figure}

In the absence of tunneling, the spectrum of the model TI described by Eq. (\ref{eq:tbmodel}) develops a bulk gap, provided an anisotropy is introduced in the hopping along the four nearest neighbor bonds, $t \rightarrow t+\delta t_{\alpha}$, $\alpha = \overline{1,4}$~\cite{Fu2007}. In the presence of a boundary, gapless surface sates develop inside the bulk gap. The results for a slab with a ($1,1,1$) surface and  parameters $t=1$, $\lambda_{\rm SO} = 0.3$ and $\delta t = 0.25$ for bonds with a surface projection oriented along M (which we choose as the $x$ direction) are shown in Fig. \ref{Fig1}. The surface states spectrum is characterized by presence of one (anisotropic) Dirac cone in the vicinity of the M point, which represents the signature of a strong topological insulator~\cite{Fu2007}.

Next, we discuss the effective model that describes the TI surface states, as a preliminary step for identifying an effective model for the TI-SC interface. Notice that, for each wave vector {\bf k} in the vicinity of M there are exactly two surface states with energies $\varepsilon_1(\bm k) \equiv \epsilon_N(\bm k)$ and $\varepsilon_2(\bm k) \equiv \epsilon_{N+1}(\bm k)$, where $\epsilon_n(\bm k)$ with $n=1, \dots ,2N$ are the energy eigenvalues for a slab consisting of $N$ layers. The corresponding wave functions are $u_{1, \bm k}(\bm r, \sigma)$ and $u_{2, \bm k}(\bm r, \sigma)$, respectively. An effective Hamiltonian describing the surface states can be obtained by projecting the full TI Hamiltonian (\ref{eq:tbmodel}) onto the subspace spanned by $u_{1, \bm k}$ and $u_{2, \bm k}$. If we choose a  basis  for this subspace defined by certain wave functions $u_{\pm, \bm k}$, an arbitrary surface state can be expressed as 
\begin{equation}
\psi_{\bm k}(\bm r, \sigma)=\sum_{\lambda=\pm} \varphi_{\lambda,\bm k}u_{\lambda,\bm k}(\bm r, \sigma), 
\end{equation}
where $\varphi_{\pm,\bm k}$ depend on the two-dimensional momentum {\bf k}. After projection, the Schrodinger equation for the surface states becomes an equation for the spinor $\phi({\bm k})=(\varphi_{+,\bm k},\varphi_{-,\bm k})^T$ or, in real space, for its Fourier transform $\phi({\bm r})$, where {\bf r} is now a 2D vector. Thus, the full three-dimensional quantum problem for the TI surface states is reduced to an effective 2D problem defined by $H_{\rm eff}(\bm k; \lambda, \lambda^{\prime}) = \langle u_{\lambda, \bm k} | H_{\rm TI} |  u_{\lambda^{\prime}, \bm k}\rangle$.  

However, choosing an appropriate basis for this subspace has to take into account two requirements that stem from  considerations of symmetry and analyticity.  First, it is convenient to have an effective Hamiltonian with manifest time-reversal (TR) properties.    
To this end,  one can  choose a pseudo-spin basis $u_{\pm, \bm k}(\bm r, \sigma)$ with the property  $\Theta u_{\pm,\bm k}=\pm u_{\mp,\bm k}$, where $\Theta =i\sigma_y K$ is TR transformation operator and K denotes the complex conjugation. Under such a choice, the spinor $\phi({\bm k})=(\varphi_{+,\bm k},\varphi_{-,\bm k})^T$ satisfies the relation $\Theta\phi({\bm k})=\imath \sigma_y \phi^*({-\bm k})$, which is identical to the time-reversal transformation of physical spinors. Given that spin is not a good quantum number in the presence of spin-orbit interactions, it is natural to construct the effective Hamiltonian using the pseudo-spin basis $u_{\mp,\bm M}$. Note that the eigenfunctions $u_{1, \bm k}$ and $u_{2, \bm k}$ have complicated TR properties, hence they  do not represent a convenient basis.  Second, we may be interested in a real space formulation of the effective theory that would apply to spatially inhomogeneous situations, for example, when impurities or vortices are present. Thus, it is desirable to construct an effective model using real space  spinors $\phi(\bm r)=(\varphi_+(\bm r), \varphi_-(\bm r))^T$. A surface state  can be expressed in terms of Wannier functions as 
\begin{align}
\psi(\mathbf{r}, \sigma)=\sum_{\lambda=\pm; \mathbf{R}}\varphi_{\lambda}(\mathbf{R})w_\lambda(\mathbf{r}-\mathbf{R},\sigma),
\end{align}
where $w_\lambda(\mathbf{r}-\mathbf{R},\sigma)=\sum_{\mathbf{k}}e^{\imath \mathbf{k}\cdot(\mathbf{r}-\mathbf{R})}u_{\lambda,\mathbf{k}}(\mathbf{r},\sigma)$ and  $u_{\pm,\bm k}$ are appropriate combinations of the energy eigenstates $u_{1,2; \bm k}$,  so that they transform as pseudo-spins under time reversal. To construct exponentially localized Wannier functions~\cite{Kohn_PR'59},  the basis states $u_{\pm \bm k}(\bm r, \sigma)$  have to be analytic in the vicinity of the M point where energy levels cross. Note that the limit $\bm k \rightarrow M$ is ill-defined for the energy eigenstates $u_{n, \bm k}$, as it depends on the direction along which the M point is approached. 

A simple way of constructing the pseudo-spin basis actually used  in our numerical analysis is to form  linear combinations of $u_{1,2; \bm k}(\bm r, \sigma)$ that are spin-polarized in the top TI layer, 
\begin{align}
\!\left(\! \begin{array}{c} \bar{u}_{+,k}(\bm r_N, \! \uparrow) \\  \bar{u}_{+,k}(\bm r_N, \! \downarrow)  \end{array} \!\right)\!=\!f(\bm k)\! \left( \!\! \begin{array}{c} 1 \\  0 \end{array} \! \right), ~~~ \left(\! \begin{array}{c} \bar{u}_{-,k}(\bm r_N, \! \uparrow) \\ \bar{u}_{-,k}(\bm r_N, \! \downarrow)  \end{array} \! \right)\!=\!f(\! - \bm k \! )\! \left(\! \begin{array}{c} 0 \\  1 \end{array} \! \right)\! ,     \label{baru}
\end{align}
 where $f(\bm k)$ is a positive smooth function. Generally, the functions $\bar{u}_{\pm, \bm k}(\bm r \sigma)$ are not orthogonal, but  one can easily construct almost spin-polarized orthogonal wave functions as 
\begin{align}
u_{+,\bm k}=N_+(\bm k)\left(\bar u_{+, \bm k}+\alpha(k) \bar u_{-, \bm k}\right), \\
u_{-,\bm k}=N_-(\bm k)\left(\bar u_{-, \bm k}+\alpha^*(k) \bar u_{+, \bm k}\right)
\end{align}     
where $\alpha(k)=-\left(1-\sqrt{1-|\langle \bar u_{+,\bm k}|  \bar u_{-,\bm k}\rangle|^2} \right)/\langle \bar u_{+,\bm k}|  \bar u_{-,\bm k}\rangle$ vanishes at the Dirac point and $N_{\pm}(k)$ are normalization factors chosen so that $u_{+,k}(\bm r_N, \! \uparrow)$ and $u_{-,k}(\bm r_N, \! \downarrow)$ be positive. The functions $u_{\pm,\bm k}$  obtained using this prescription have the  transformation properties of spinors under time reversal and are  smooth in $\bm k$ around the M point. The construction is valid for all values of $\bm k$ corresponding to surface modes.

If we are only interested in the properties of the surface states in the vicinity of the M point, the effective 2D model can be constructed using the pseudo-spin basis at the M point, $u_{\pm,\bm k_M}$,  and $\bm k \cdot \bm p$ perturbation theory. One obtains 
\begin{align}\label{eq:Hamiltonian2}
H_{TI}^{\rm eff} \approx E_M+\bm B(k)\cdot \bm \sigma,
\end{align}
where $E_M$ is the energy at the M point, $\bm B(\bm k)$ is a linear function of  $\bm k$ and $\bm \sigma=(\sigma_x,\sigma_y,\sigma_z)$ are Pauli matrices representing the pseudo-spin. The most general form for the effective Hamiltonian  can be determined  using the time-reversal properties of the pseudo-spin basis $u_{\pm,\bm k}$:
\begin{equation}
H_{TI}^{\rm eff}=L_0(\mathbf{k})+\sum_{i=1,2,3}L_{i}(\mathbf{k})\sigma_i   \label{Heff}
\end{equation}
where $L_0(\mathbf{k})$ and $L_i(\mathbf{k})$ are even and odd smooth functions of $\mathbf{k}$, respectively. Unlike, the Hamiltonian \eqref{eq:Hamiltonian2} derived from $k\cdot p$ perturbation theory, the above Hamiltonian is valid for all momenta over which the surface states are defined. For a generic anisotropic TI model  an important feature is the presence $L_z(\bm k)$, which leads to unequal spin population.  Note that in the limit of $k\rightarrow M$ we recover the Hamiltonian \eqref{eq:Hamiltonian2}. 

\begin{figure}[tbp]
\begin{center}
\includegraphics[width=0.47\textwidth]{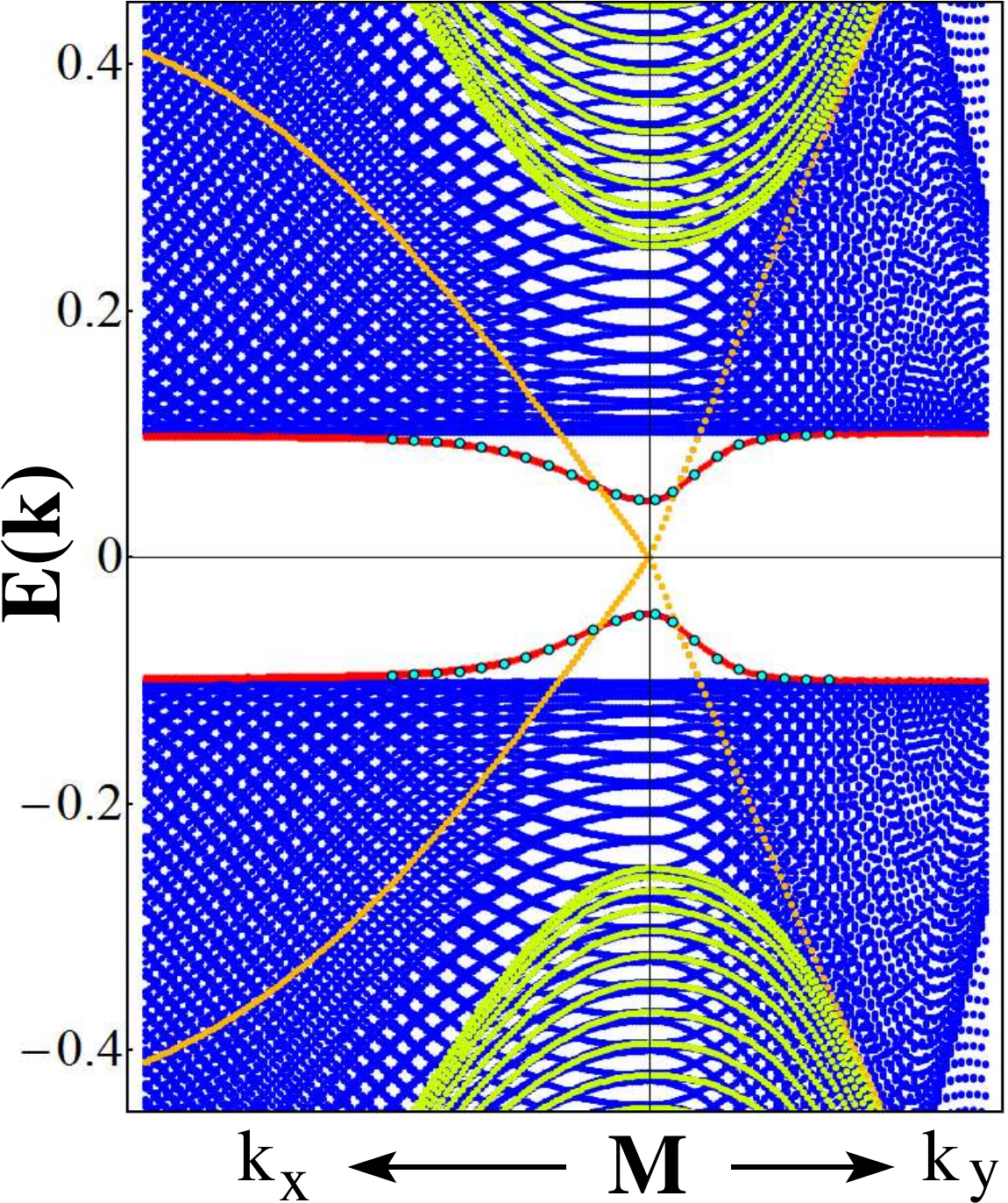}
\end{center}
\caption{(Color online) Spectrum of the full Hamiltonian (\ref{Htot}) for a TI slab with (1,~1,~1) surfaces having an interface with an $s$-wave superconductor. Bulk TI states are green (light gray) and SC states are blue (dark gray areas). The proximity effect induces a gap in the surface states spectrum at the interface (red/dark gray in-gap lines), while the states on the opposite surface have an unperturbed Dirac cone dispersion (orange/light gray lines). The light blue (light gray) in-gap points are calculated using an effective theory for the interface described by Eqs. (\ref{Heff}), and (\ref{FN}-\ref{eq:spectrum}).} \vspace{-4mm}
\label{Fig2}
\end{figure}

Next, we turn on the coupling between the TI and the SC. As a result of the proximity effect, a finite gap opens in the surface states spectrum. The result of the numerical calculation for the full tight-binding Hamiltonian (\ref{Htot}) is shown in Fig. \ref{Fig2} for the same TI model  parameters as in Fig. \ref{Fig1}, a superconductor with $t_{||}^{\prime}=0.5t$ and $t_{\perp}^{\prime} =0.45$ for hoppings parallel  and perpendicular to the interface, respectively, and $\Delta_0 = 0.1t$ and an interface characterized by a transparency $\tilde{t} = 0.3t$. To better understand the dependence of the proximity effect on various model parameters, it is useful  to identify the SC contribution to the interface effective model and add it to TI contribution given by Eq. (\ref{Heff}).
Within the mean-field model  given by Eq.~\eqref{eq:scmodel}, the superconducting degrees of freedom can be integrated out and replaced by the interface  self-energy $\Sigma_{\sigma \sigma}(\bm{r}, \bm r', \omega)=-\tilde T (\bm r, \bm r_1)\hat G_{\sigma \sigma'}(\bm r_1, \bm r_2; \omega)\tilde T^\dagger(\bm r_2, \bm r')$, 
where $\hat{G}_{\sigma \sigma'} (\bm r_1, \bm r_2; \omega)$ is the superconducting Green's function in the Nambu space and $\tilde T$ is the matrix describing the tunneling between top layer of the TI (layer $N$ of the TI-SC heterostructure) and bottom layer of the superconductor (layer N+1). For an $s$-wave superconductor, we have 
\begin{align}
\Sigma(\bm k, \omega)=-\tilde t^2 \sum_m \frac{\omega \tau_0 +\epsilon_m(\bm k)\tau_z +\Delta_0 \tau_x}{\epsilon_m(\bm k)^2+\Delta_0^2-\omega^2}|\chi_{m \bm k}(z_{N+1})|^2   \label{SelfE2}
\end{align}
with $\chi_{m \bm k}(z_{N+1})$ being the value on the bottom layer of the energy eigenstates in the metallic phase. This self-energy in terms of the partial density of states at the interface, $ \nu(\varepsilon, \bm k)=\sum_{m}\delta(\varepsilon - \epsilon_m(\bm k))|\chi_{m \bm k}(z_{N+1})|^2 $, which in our model takes the simple form $ \nu(\varepsilon, \bm k)=
\frac{2}{\pi\Lambda}\sqrt{1-\frac{[\varepsilon-\bar{\epsilon}(\bm k)]^2}{\Lambda^2}}$, where $\Lambda = 2t_{\perp}$ is the bandwidth  in the metallic phase of a slab described by the Hamiltonian  (\ref{eq:scmodel}) and $\bar{\epsilon}(\bm k)$ is the energy value at the middle of the band for a given momentum parallel to the interface. For surface states in the vicinity of $\bm k\approx \bm k_M$ and assuming $\Delta_0\ll\Lambda$, we have 
\begin{equation}
\Sigma(\bm k, \omega) \approx -\lambda \frac{ (\omega \tau_0 +\Delta_0\tau_x)}{\sqrt{\Delta_0^2-\omega^2}},   \label{SelfE3}
\end{equation}
where $\lambda = 2\tilde{t}^2/\Lambda\sqrt{1-(\bar{\epsilon}(\bm k_M)-\mu)^2/\Lambda^2}$, with $\mu$ being the chemical potential of the metal.
Note that this approximation ignores the momentum dependence of the self-energy induced by the $s$-wave superconductor.

To construct the effective model for the proximity effect, we project the above equation onto pseudo-spin basis $u_{\pm,\bm k} (\bm r, \sigma)$. The effective self-energy at the interface becomes
\begin{align}
F^{(N)}_{\lambda \lambda'}(\bm k, \omega)&=\sum_{\sigma \sigma'} u^*_{\lambda \bm k}(\bm r_N, \sigma) \Sigma_{\sigma, \sigma'}^{(N)}(\bm k, \omega) u_{\lambda', \bm k} (\bm r_N, \sigma') \nonumber \\
&\approx \frac{ \lambda \omega}{\sqrt{\Delta_0^2-\omega^2} } \sum_{\sigma} u^*_{\lambda \bm k}(\bm r_N, \sigma) u_{\lambda', \bm k} (\bm r_N, \sigma) \label{FN}\\
F^{(A)}_{\lambda \lambda'}(\bm k, \omega)&=\sum_{\sigma \sigma'}  u_{\lambda \bm k}(\bm r_N, \sigma) \Sigma_{\sigma, \sigma'}^{(A)}(\bm k, \omega) u_{\lambda', - \bm k} (\bm r_N, \sigma') \nonumber \\
&\approx \frac{ \lambda \Delta_0}{\sqrt{\Delta_0^2-\omega^2} } \sum_{\sigma} u_{\lambda \bm k}(\bm r_N, \sigma) u_{\lambda', -\bm k} (\bm r_N, -\sigma) .  \label{FA}
\end{align}
Here $\Sigma^{N/A}$ are the normal (diagonal) and anomalous (off-diagonal) parts of the self-energy and $\bm k=(k_x,k_y)$.  In general, both $F_{\lambda \lambda'}^{(N)}$ and $F_{\lambda \lambda'}^{(A)}$ have even and odd components in $\bm k$.  The anomalous part of $F^{(A)}_{\lambda,\lambda'}(\bm k, \omega)$ is an even function of frequency and has singlet $\Delta_{s}(\mathbf{k},\omega)=\Delta_{s}(-\mathbf{k},\omega)$ and triplet $\Delta_{t}(\mathbf{k},\omega)=-\Delta_{t}(-\mathbf{k},\omega)$ components.  
Note that the relevant $k$-dependent corrections arise from the pseudo-spin eigenstates $u_{\lambda, \bm k} (\bm r_N,\sigma)$. Even though the construction of the pseudo-spin basis is defined up to a $k$-dependent unitary transformation that preserve the time-reversal properties of the pseudo-spin basis, such transformation preserves $|\Delta_{s}(\mathbf{k},\omega)|^2$ and $|\Delta_{t}(\mathbf{k},\omega)|^2$. We emphasize that the emergence of the triplet component can also be viewed as a property of the ${\mathbf k}\cdot{\mathbf p}$ perturbation theory for the Bogoliubov-de Gennes (BdG) Hamiltonian. This property is generic and independent of the position of the Dirac point.
Finally,  the spectrum of excitations at the interface can be determined by solving the BdG equation 
\begin{align}
{\rm Det} (H_{\rm TI}^{\rm eff}(\bm k)+F(\bm k, \omega)-\omega)=0. \label{eq:spectrum}
\end{align}

\begin{figure}[tbp]
\begin{center}
\includegraphics[width=0.45\textwidth]{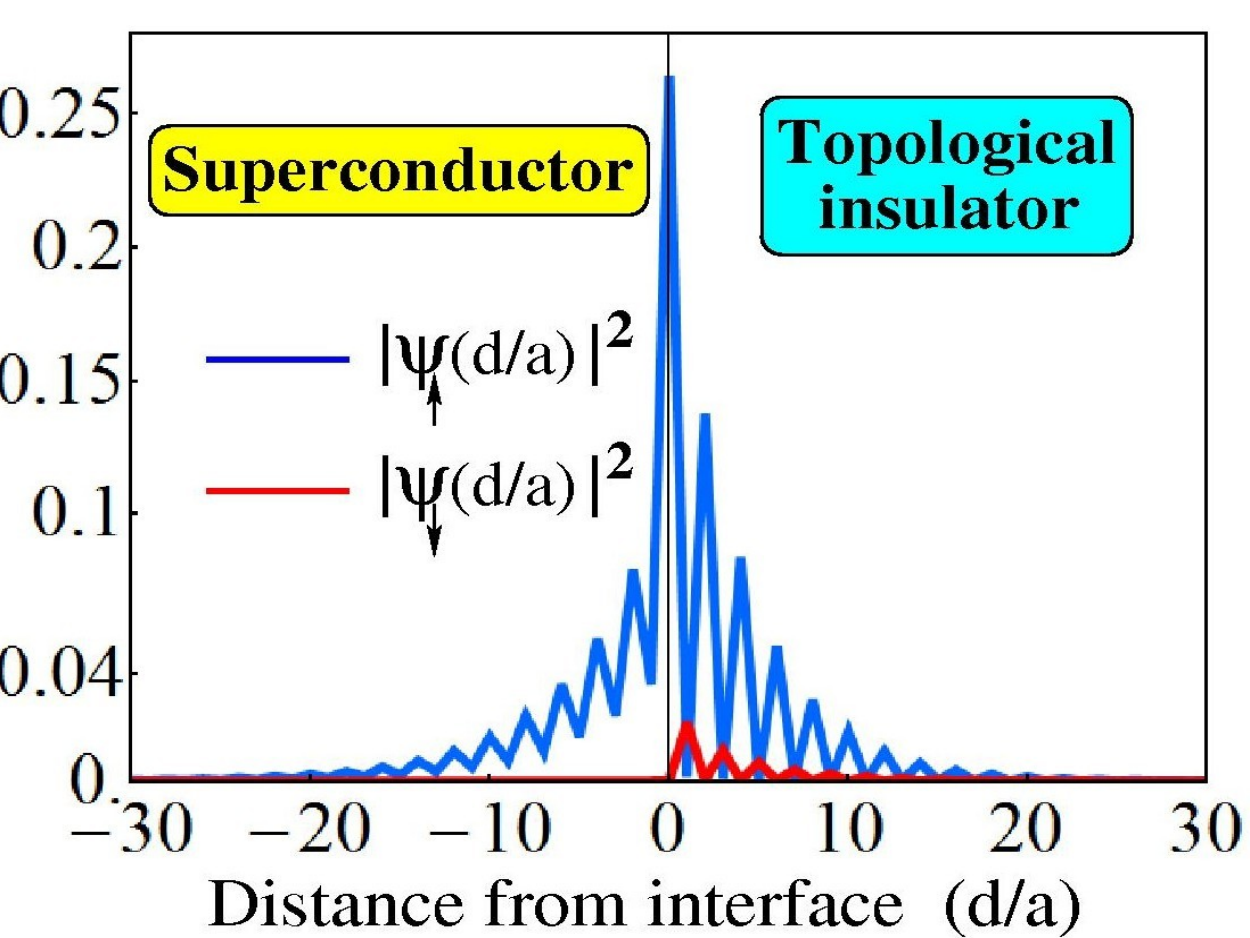}
\end{center}
\caption{(Color online) Surface state localized at the interface between a TI and an $s$-wave SC. For zero tunneling (not shown), the state resides entirely inside the TI and decays exponentially  away from the boundary (up to an even-odd oscillatory factor). At finite transparency, the surface state penetrates inside the SC. The spin-down (lower amplitude) component vanishes on the top TI layer, thus it does not propagate inside the SC if only nearest-neighbor tunneling is considered.}
\label{Fig3} \vspace{-4mm}
\end{figure}

A comparison between the spectrum of the TI-SC heterostructure described by Eq. (\ref{Htot}) and the spectrum of the effective model in Eq. (\ref{eq:spectrum}) is shown in Fig. \ref{Fig2} for a system with a chemical potential that crosses the Dirac point, $\mu=0$. Similar calculations were performed for systems with $\mu\neq 0$ and for systems with a different interface orientation, i.e., (100) instead of (111). Adding a potential barrier at the interface was also considered but the results remain qualitatively the same. 

We now discuss the main implications of the microscopic calculations described above.  In the limit $k\approx k_M$ and $\mu\approx 0$ the triplet
 component $\Delta_t$ vanishes due to the time-reversal symmetry and  pairing becomes purely $s$-wave, as conjectured in Ref.~\cite{Fu2008}. The low-energy spectrum can be obtained by expanding the self-energy to lowest order in $\bm k$ and $\omega$
\begin{equation}
{\rm Det} \left[\frac{1}{1+\lambda/\Delta}(v \bm \sigma \bm k)\tau_z+\Delta\frac{\lambda}{\lambda+\Delta}\tau_x-\omega \right]=0.
\end{equation}
We find that in this limit, consistent with previous phenomenological
results~\cite{sau'09}, the form of the above effective
 BdG equation resembles the equation for the proximity effect
 at the interface of a TI and an $s$-wave superconductor\cite{Fu2008}
 with the critical difference that  the effective parameters of the TI
 surface states such as the velocity $v$ and chemical potential $\mu$ are
 strongly renormalized by $1+\lambda/\Delta$. 
 This renormalization, which can be thought of as a consequence of the 
penetration of the TI wave function into the SC (see Fig.~\ref{Fig3}), determines the energy scale for all subgap excitations at the interface, e.g., the Majorana modes. In particular, it determines the size of the minigap that  protects the Majorana bound state, hence it has direct consequences for
 the robustness of topological quantum computation~\cite{sau'09}.

As the chemical potential is tuned away from the Dirac point $\mu \approx 0$ the triplet component of the pairing potential develops. This follows from the fact that the wave functions on the TI are not
spin-polarized due to the spin-orbit coupling as shown in Fig.~\ref{Fig3}. 
The nature of the pairing away from the Dirac point can be analyzed by considering the BdG
Eq.~(\ref{eq:spectrum}), for $\omega,\lambda\ll\Delta_0$ (to
ignore retardation effects for simplicity), and keeping
the lowest-order $\bm k$ dependence as ${\rm Det}\left([\mathbf{v}_i \cdot\mathbf{k}\sigma_i\!-\!\mu]\tau_z+\lambda[1\!+\!\mathbf{u}_i \cdot \bm  k\sigma_i]\tau_x\right)=0$.
Thus in addition to the proximity-induced singlet pairing of magnitude
 $\lambda$, there is a proximity induced spin-triplet pairing of magnitude
 $\lambda|u|k_F$ where $k_F\sim \mu/v$ is the
Fermi wave-vector on the Dirac cone. This triplet pairing can become
significant when $\mu \gtrsim v \lambda/u$ and can lead to a closing of the superconducting gap on the 
surface state of the TI, which has important consequences for topological quantum computation. In particular, the resulting proximity induced superconducting
state can be qualitatively different from the one where the chemical
potential is near the Dirac point and is not expected to
support Majorana fermions.

We calculated the excitation spectrum of a topological insulator in contact with an $s$-wave superconductor, starting from a microscopic model, and we developed an effective low-energy model for the proximity effect. We found that a $p$-wave pairing component generally leads to the reduction in the induced quasiparticle gap. We also showed that virtual propagation of the electrons into the superconductor leads to a significant renormalization of the effective surface TI Hamiltonian. These findings have important implications for topological quantum computation.

This work is supported by DARPA-QuEST, JQI-PFC and LPS-NSA.


\vspace{-4mm}

\begin{thebibliography}{24}
\expandafter\ifx\csname natexlab\endcsname\relax\def\natexlab#1{#1}\fi
\expandafter\ifx\csname bibnamefont\endcsname\relax
  \def\bibnamefont#1{#1}\fi
\expandafter\ifx\csname bibfnamefont\endcsname\relax
  \def\bibfnamefont#1{#1}\fi
\expandafter\ifx\csname citenamefont\endcsname\relax
  \def\citenamefont#1{#1}\fi
\expandafter\ifx\csname url\endcsname\relax
  \def\url#1{\texttt{#1}}\fi
\expandafter\ifx\csname urlprefix\endcsname\relax\def\urlprefix{URL }\fi
\providecommand{\bibinfo}[2]{#2}
\providecommand{\eprint}[2][]{\url{#2}}

\bibitem[{\citenamefont{Konig et~al.}(2007)\citenamefont{Konig, Wiedmann,
  Brune, Roth, Buhmann, Molenkamp, Qi, and Zhang}}]{Konig2007}
\bibinfo{author}{\bibfnamefont{M.}~\bibnamefont{Konig}},
\bibinfo{author}{\bibfnamefont{S.}~\bibnamefont{Wiedmann}},
\bibinfo{author}{\bibfnamefont{C.}~\bibnamefont{Brune}},
\bibinfo{author}{\bibfnamefont{A.}~\bibnamefont{Roth}},
\bibinfo{author}{\bibfnamefont{H.}~\bibnamefont{Buhmann}},
\bibinfo{author}{\bibfnamefont{L.~W.} \bibnamefont{Molenkamp}},
\bibinfo{author}{\bibfnamefont{X.-L.} \bibnamefont{Qi}}, \bibnamefont{and}
\bibinfo{author}{\bibfnamefont{S.-C.} \bibnamefont{Zhang}},
  \bibinfo{journal}{Science} \textbf{\bibinfo{volume}{318}},
  \bibinfo{pages}{766} (\bibinfo{year}{2007}).

\bibitem[{\citenamefont{Hsieh et~al.}(2008)\citenamefont{Hsieh, Qian, Wray,
  Xia, Hor, Cava, and Hasan}}]{Hsieh2008}
\bibinfo{author}{\bibfnamefont{D.}~\bibnamefont{Hsieh}},
\bibinfo{author}{\bibfnamefont{D.}~\bibnamefont{Qian}},
\bibinfo{author}{\bibfnamefont{L.}~\bibnamefont{Wray}},
\bibinfo{author}{\bibfnamefont{Y.}~\bibnamefont{Xia}},
\bibinfo{author}{\bibfnamefont{Y.~S.} \bibnamefont{Hor}},
\bibinfo{author}{\bibfnamefont{R.~J.} \bibnamefont{Cava}}, \bibnamefont{and}
\bibinfo{author}{\bibfnamefont{M.~Z.} \bibnamefont{Hasan}},
  \bibinfo{journal}{Nature (London)} \textbf{\bibinfo{volume}{452}},
  \bibinfo{pages}{970} (\bibinfo{year}{2008}).

\bibitem[{\citenamefont{Hsieh et~al.}(2009)\citenamefont{Hsieh, Xia, Wray,
  Qian, Pal, Dil, Osterwalder, Meier, Bihlmayer, and Kane}}]{Hsieh2009}
\bibinfo{author}{\bibfnamefont{D.}~\bibnamefont{Hsieh}},
\bibinfo{author}{\bibfnamefont{Y.}~\bibnamefont{Xia}},
\bibinfo{author}{\bibfnamefont{L.}~\bibnamefont{Wray}},
\bibinfo{author}{\bibfnamefont{D.}~\bibnamefont{Qian}},
\bibinfo{author}{\bibfnamefont{A.}~\bibnamefont{Pal}},
\bibinfo{author}{\bibfnamefont{J.~H.} \bibnamefont{Dil}},
\bibinfo{author}{\bibfnamefont{J.}~\bibnamefont{Osterwalder}},
\bibinfo{author}{\bibfnamefont{F.}~\bibnamefont{Meier}},
\bibinfo{author}{\bibfnamefont{G.}~\bibnamefont{Bihlmayer}},\bibnamefont{and}
\bibinfo{author}{\bibfnamefont{C.~L.} \bibnamefont{Kane}}
  \bibinfo{journal}{Science}
  \textbf{\bibinfo{volume}{323}}, \bibinfo{pages}{919} (\bibinfo{year}{2009}).

\bibitem[{\citenamefont{Roushan et~al.}(2009)\citenamefont{Roushan, Seo,
  Parker, Hor, Hsieh, Qian, Richardella, Hasan, Cava, and
  Yazdani}}]{Roushan2009}
\bibinfo{author}{\bibfnamefont{P.}~\bibnamefont{Roushan}},
\bibinfo{author}{\bibfnamefont{J.}~\bibnamefont{Seo}},
\bibinfo{author}{\bibfnamefont{C.~V.}~\bibnamefont{Parker}},
\bibinfo{author}{\bibfnamefont{Y.~S.}~\bibnamefont{Hor}},
\bibinfo{author}{\bibfnamefont{D.}~\bibnamefont{Hsieh}},
\bibinfo{author}{\bibfnamefont{D.}~\bibnamefont{Qian}},
\bibinfo{author}{\bibfnamefont{A.}~\bibnamefont{Richardella}},
\bibinfo{author}{\bibfnamefont{M.~Z.}~\bibnamefont{Hasan}},
\bibinfo{author}{\bibfnamefont{R.~J.}~\bibnamefont{Cava}},\bibnamefont{and}
\bibinfo{author}{\bibfnamefont{A.}~\bibnamefont{Yazdani}},
  \bibinfo{journal}{Nature} \textbf{\bibinfo{volume}{460}},
  \bibinfo{pages}{1106} (\bibinfo{year}{2009}).

\bibitem[{\citenamefont{Zhang et~al.}(2009)\citenamefont{Zhang, Cheng, Chen,
  Jia, Ma, He, Wang, Zhang, Dai, Fang et~al.}}]{TZhang2009}
\bibinfo{author}{\bibfnamefont{T.}~\bibnamefont{Zhang  {\it et al.}}},
  \bibinfo{journal}{Phys. Rev. Lett.} \textbf{\bibinfo{volume}{103}},
  \bibinfo{pages}{266803} (\bibinfo{year}{2009}).


\bibitem[{\citenamefont{Y. Xia et~al.}(2009)\citenamefont{Xia, Qian,
  Hsieh, Wray, Pal, Lin, Bansil, Grauer, Hor, Cava, and Hasan}}]{Xia2009}
\bibinfo{author}{\bibfnamefont{Y.}~\bibnamefont{Xia}},
\bibinfo{author}{\bibfnamefont{D.}~\bibnamefont{Qian}},
\bibinfo{author}{\bibfnamefont{D.}~\bibnamefont{Hsieh}},
\bibinfo{author}{\bibfnamefont{L.}~\bibnamefont{Wray}},
\bibinfo{author}{\bibfnamefont{A.}~\bibnamefont{Pal}},
\bibinfo{author}{\bibfnamefont{H.}~\bibnamefont{Lin}},
\bibinfo{author}{\bibfnamefont{A.}~\bibnamefont{Bansil}},
\bibinfo{author}{\bibfnamefont{D.}~\bibnamefont{Grauer}},
\bibinfo{author}{\bibfnamefont{Y.~S.}~\bibnamefont{Hor}},
\bibinfo{author}{\bibfnamefont{R.~J.}~\bibnamefont{Cava}},\bibnamefont{and}
\bibinfo{author}{\bibfnamefont{M.~Z.}~\bibnamefont{Hasan}},
  \bibinfo{journal}{Nat. Phys.} \textbf{\bibinfo{volume}{5}},
  \bibinfo{pages}{398} (\bibinfo{year}{2009}).

\bibitem[{\citenamefont{Kitaev}(2003)}]{Kitaev2003}
\bibinfo{author}{\bibfnamefont{A.~Y.} \bibnamefont{Kitaev}},
  \bibinfo{journal}{Ann. Phys. (N.Y.)} \textbf{\bibinfo{volume}{303}},
  \bibinfo{pages}{2} (\bibinfo{year}{2003}).

\bibitem[{\citenamefont{Kane and Mele}(2005)}]{KaneMele2005}
\bibinfo{author}{\bibfnamefont{C.~L.} \bibnamefont{Kane}} \bibnamefont{and}
  \bibinfo{author}{\bibfnamefont{E.~J.} \bibnamefont{Mele}},
  \bibinfo{journal}{Phys. Rev. Lett.} \textbf{\bibinfo{volume}{95}},
  \bibinfo{pages}{146802} (\bibinfo{year}{2005}).

\bibitem[{\citenamefont{Bernevig and Zhang}(2006)}]{Bernevig2006}
\bibinfo{author}{\bibfnamefont{B.~A.} \bibnamefont{Bernevig}} \bibnamefont{and}
  \bibinfo{author}{\bibfnamefont{S.-C.} \bibnamefont{Zhang}},
  \bibinfo{journal}{Phys. Rev. Lett.} \textbf{\bibinfo{volume}{96}},
  \bibinfo{pages}{106802} (\bibinfo{year}{2006}).

\bibitem[{\citenamefont{Murakami}(2006)}]{Murakami2006}
\bibinfo{author}{\bibfnamefont{S.}~\bibnamefont{Murakami}},
  \bibinfo{journal}{Phys. Rev. Lett.} \textbf{\bibinfo{volume}{97}},
  \bibinfo{pages}{236805} (\bibinfo{year}{2006}).

\bibitem[{\citenamefont{Wu et~al.}(2006)\citenamefont{Wu, Bernevig, and
  Zhang}}]{Wu2006}
\bibinfo{author}{\bibfnamefont{C.}~\bibnamefont{Wu}},
  \bibinfo{author}{\bibfnamefont{B.~A.} \bibnamefont{Bernevig}},
  \bibnamefont{and} \bibinfo{author}{\bibfnamefont{S.-C.} \bibnamefont{Zhang}},
  \bibinfo{journal}{Phys. Rev. Lett.} \textbf{\bibinfo{volume}{96}},
  \bibinfo{pages}{106401} (\bibinfo{year}{2006}).

\bibitem[{\citenamefont{Fu and Kane}(2006)}]{Fu2006}
\bibinfo{author}{\bibfnamefont{L.}~\bibnamefont{Fu}} \bibnamefont{and}
  \bibinfo{author}{\bibfnamefont{C.~L.} \bibnamefont{Kane}},
  \bibinfo{journal}{Phys. Rev. B }
  \textbf{\bibinfo{volume}{74}}, \bibinfo{pages}{195312}
  (\bibinfo{year}{2006}).

\bibitem[{\citenamefont{Fu et~al.}(2007)\citenamefont{Fu, Kane, and
  Mele}}]{Fu2007}
\bibinfo{author}{\bibfnamefont{L.}~\bibnamefont{Fu}},
  \bibinfo{author}{\bibfnamefont{C.~L.} \bibnamefont{Kane}}, \bibnamefont{and}
  \bibinfo{author}{\bibfnamefont{E.~J.} \bibnamefont{Mele}},
  \bibinfo{journal}{Phys.\ Rev.\ Lett.} \textbf{\bibinfo{volume}{98}},
  \bibinfo{pages}{106803} (\bibinfo{year}{2007}).

\bibitem[{\citenamefont{Fu and Kane}(2008)}]{Fu2008}
\bibinfo{author}{\bibfnamefont{L.}~\bibnamefont{Fu}} \bibnamefont{and}
  \bibinfo{author}{\bibfnamefont{C.~L.} \bibnamefont{Kane}},
  \bibinfo{journal}{Phys.\ Rev.\ Lett.} \textbf{\bibinfo{volume}{100}},
  \bibinfo{pages}{096407} (\bibinfo{year}{2008}).

\bibitem[{\citenamefont{Heersche et~al.}(2007)\citenamefont{Heersche,
  Jarillo-Herrero, Oostinga, Vandersypen, and Morpurgo}}]{Heersche'07}
\bibinfo{author}{\bibfnamefont{H.~B.} \bibnamefont{Heersche}}, 
\bibinfo{author}{\bibfnamefont{P.} \bibnamefont{Jarillo-Herrero}}, 
\bibinfo{author}{\bibfnamefont{J.~B.} \bibnamefont{Oostinga}}, 
\bibinfo{author}{\bibfnamefont{L.~M.~K.} \bibnamefont{Vandersypen}}, \bibnamefont{and}
\bibinfo{author}{\bibfnamefont{A.~F.} \bibnamefont{Morpurgo}}, 
\bibinfo{journal}{Nature}
  \textbf{\bibinfo{volume}{446}}, \bibinfo{pages}{56} (\bibinfo{year}{2007}).

\bibitem[{\citenamefont{Miao et~al.}(2009)\citenamefont{Miao, Bao, Zhang, and
  Lau}}]{Miao'09}
\bibinfo{author}{\bibfnamefont{F.}~\bibnamefont{Miao}},
\bibinfo{author}{\bibfnamefont{W.}~\bibnamefont{Bao}},
\bibinfo{author}{\bibfnamefont{H.}~\bibnamefont{Zhang}}, \bibnamefont{and}
\bibinfo{author}{\bibfnamefont{C.~N.} \bibnamefont{Lau}},
  \bibinfo{journal}{Solid State Commun.} \textbf{\bibinfo{volume}{149}},
  \bibinfo{pages}{1046 } (\bibinfo{year}{2009}).

\bibitem[{\citenamefont{Lutchyn et~al.}(2008)\citenamefont{Lutchyn, Galitski,
  Refael, and Das~Sarma}}]{lutchyn'08}
\bibinfo{author}{\bibfnamefont{R.~M.} \bibnamefont{Lutchyn}},
\bibinfo{author}{\bibfnamefont{V.}~\bibnamefont{Galitski}},
\bibinfo{author}{\bibfnamefont{G.}~\bibnamefont{Refael}}, \bibnamefont{and}
\bibinfo{author}{\bibfnamefont{S.}~\bibnamefont{Das~Sarma}},
  \bibinfo{journal}{Phys. Rev. Lett.} \textbf{\bibinfo{volume}{101}},
  \bibinfo{pages}{106402} (\bibinfo{year}{2008}).

\bibitem[{\citenamefont{Kohn}(1959)}]{Kohn_PR'59}
\bibinfo{author}{\bibfnamefont{W.}~\bibnamefont{Kohn}}, \bibinfo{journal}{Phys.
  Rev.} \textbf{\bibinfo{volume}{115}}, \bibinfo{pages}{809}
  (\bibinfo{year}{1959}).

\bibitem[{\citenamefont{J.D. Sau et~al.}(2009)\citenamefont{Sau, Lutchyn, Tewari, Das Sarma}}]{sau'09}
\bibinfo{author}{\bibnamefont{J.~D. Sau}},
\bibinfo{author}{\bibfnamefont{R.~M.} \bibnamefont{Lutchyn}},
\bibinfo{author}{\bibfnamefont{S} \bibnamefont{Tewari}}, \bibnamefont{and}
\bibinfo{author}{\bibfnamefont{S.}~\bibnamefont{Das~Sarma}},
  \bibinfo{journal}{Phys. Rev. Lett.} \textbf{\bibinfo{volume}{104}},
  \bibinfo{pages}{040502} (\bibinfo{year}{2010}).

\end{thebibliography}
\end{document}